\documentstyle[12pt,aasms4]{article}

\def\arcsecpoint{$''\!.$}

\lefthead{Kraemer \& Crenshaw}
\righthead{Resolved Spectroscopy of the Narrow-Line Region in NGC 1068.
 II, Physical Conditions Near the NGC 1068 ``Hot-Spot''}

\begin{document}

\title{Resolved Spectroscopy of the Narrow-Line Region in NGC 1068.\\
II. Physical Conditions Near the NGC 1068 ``Hot-Spot''\altaffilmark{1}}

\author{Steven B. Kraemer\altaffilmark{2,3},
\& D. Michael Crenshaw\altaffilmark{2,4}}

\altaffiltext{1}{Based on observations made with the NASA/ESA Hubble Space 
Telescope. STScI is operated by the Association of Universities for Research in 
Astronomy, Inc. under the NASA contract NAS5-26555. }

\altaffiltext{2}{Catholic University of America,
NASA/Goddard Space Flight Center, Code 681,
Greenbelt, MD  20771.}

\altaffiltext{3}{Email: stiskraemer@yancey.gsfc.nasa.gov.}

\altaffiltext{4}{Email: crenshaw@buckeye.gsfc.nasa.gov.}

\begin{abstract}

  The physical conditions near the optical continuum peak 
  (``hot spot'') in the inner narrow 
  line region (NLR) of the Seyfert
  2 galaxy, NGC 1068, are examined using ultraviolet and optical spectra 
  and photoionization models. The spectra were taken with the
  {\it Hubble Space Telescope}/Space Telescope Imaging Spectrograph 
  ({\it HST}/STIS),
  through the 0\arcsecpoint1X52\arcsecpoint0 slit, covering the 
  full STIS 1200 \AA\ to 10000 \AA\ waveband, and
  are from a region that includes the
  hot spot, extending 0\arcsecpoint2, or $\sim$ 14 pc 
  (for H$_{0}$ $=$ 75 km sec$^{-1}$ Mpc$^{-1}$), in the cross-dispersion
  direction.
  The spectra show emission-lines from a wide range 
  of ionization states for the most abundant elements, similar to 
  archival Faint Object Spectrograph spectra of the same region. 
  Perhaps the most striking feature of these spectra is the presence of 
  strong coronal emission lines, including [S~XII] $\lambda$7611 which
  has hitherto only been identified in spectra of the solar corona.
  There is an apparent correlation between ionization energy
  and velocity of the emission lines with respect to the systemic velocity of the
  host galaxy, with the coronal lines blueshifted, most other high excitation 
  lines near systemic, and some of the low ionization lines redshifted.
  From the results of our modeling, we find that the 
  emission-line
  gas is photoionized and consists of three principal components:
  1) one in which most of the strong emission-lines, such
  as [O~III] $\lambda$5007, [Ne~V] $\lambda$3426, C~IV $\lambda$1550,
  arise, 2) a more tenuous, highly ionized component, which is the source of the
  coronal-line emission, and 3) a component, which is not co-planar with
  the other two, in which the low ionization and neutral lines, such as
  [N~II] $\lambda$6548 and [O~I] $\lambda$6300, are formed. The first
  two components are directly ionized by the EUV-Xray continuum emitted
  by the central source, while the low ionization gas is ionized by
  a combination of highly absorbed continuum radiation and a small
  fraction of unabsorbed continuum scattered by free electrons associated
  with the hot spot. The combination of covering
  factor and Thomson optical depth of the high ionization components
  is insufficient to scatter the observed fraction of continuum radiation
  into our line-of-sight. Therefore, the scattering must occur in an 
  additional component of hot plasma, which contributes little or no
  UV/optical line emission.

\end{abstract}

\keywords{galaxies: individual (NGC 1068) -- galaxies: Seyfert}

\section{Introduction}

   NGC 1068, one of the initial set of emission line galaxies studied
   by Seyfert (1943), is the nearest (z=0.0038) and best studied 
   of the Seyfert 2 galaxies. Based on the widths of their emission lines
   in optical spectra, Seyfert galaxies are generally divided into
   two types (Khachikian \& Weedman 1971). Seyfert 1s possess broad permitted 
   lines, with full widths at half maximum (FWHM) 
   $\geq$ 10$^{3}$ km s$^{-1}$, and narrower forbidden lines, with 
   FWHM $\approx$ 500 km s$^{-1}$, while
   Seyfert 2s show only narrow emission lines. The optical continua
   of Seyfert 1 galaxies are dominated by nonstellar emission that
   can be characterized by a power-law (Oke \& Sargent 1968), whereas
   this component is much weaker in Seyfert 2 galaxies (Koski 1978).
   Spectropolarimetric studies 
   (Miller \& Antonucci 1983; Antonucci \& Miller 1985) revealed the
   presence of strongly polarized continuum and broad permitted line
   emission within a 3\arcsecpoint0 aperture centered on the optical
   nucleus of NGC 1068. These observations were the inspiration for the
   unified model for Seyfert galaxies, according to which the differences
   between types 1 and 2 are due to viewing angle, with Seyfert 2
   galaxies characterized by obscuration of their broad-line regions
   and central engines (cf. Antonucci 1994).
   The polarization is wavelength 
   independent, which
   implies that the emission from 
   the hidden continuum source and broad-line region is scattered into
   our line-of-sight by free electrons within a hot ($\sim$ 3 x 10$^{5}$ K)
   plasma in the inner Narrow-Line Region (NLR) (Miller, Goodrich, \&
   Mathews 1991).

   The structure of the nuclear region has been observed extensively
   with {\it HST}, using the Planetary Camera (pre-COSTAR; Lynds et al. 
   1991), and the Faint Object Camera (Capetti et al. 1995). The
   peak in the optical continuum is resolved, with a FWHM of approximately 
   0\arcsecpoint15, roughly centered within a cloud of 
   dimensions 3\arcsecpoint5 x 1\arcsecpoint7 elongated
   in the NE-SW direction, the latter consisting primarily of 
   starlight (Lynds et al. 1991;
   Crenshaw \& Kraemer 1999, hereafter Paper I). 
   The NE part of the cloud has the physical dimensions
   and location attributed to the scattering medium modeled by Miller,
   et al. (1991), which has been revealed in the
   UV (Kriss et al. 1993; Macchetto et al. 1994) and which we have discussed in detail in
   Paper I.  A bright knot of [O~III] $\lambda$5007 emission, often 
   referred to as Cloud B (Evans et al. 1991), overlaps the continuum
   emission from the hot spot; its centroid is $\sim$ 0\arcsecpoint2
   north of that of the hot spot.

   While Cloud B lies close to the apex of emission-line bicone (Evans et al.
   1991), there is evidence that the actual hidden nucleus is further
   to the south and west (Capetti, Macchetto, \& Lattanzi 1997). It has
   been suggested that a thermal radio source, S1, 
   0\arcsecpoint3 south of the continuum peak, is the true
   nucleus (Evans et al. 1991; Gallimore et al 1997). The discovery of an H$_{2}$O megamaser
   with a velocity width of 600 km s$^{-1}$ (Claussen \& Lo 1986)
   associated with S1 (cf. Greenhill \& Gwinn 1997), is further evidence
   that S1 is the nucleus. Even so, the optical continuum peak and
   a portion of Cloud B are within 30 pc of the nucleus, and therefore exposed 
   to an intense flux of ionizing radiation.

   Coronal iron lines have been detected in a number of Seyfert 
   galaxies (Grandi 1978; Osterbrock 1981; Penston et al. 1984; Osterbrock
   1985). The following coronal-lines are known to be present 
   in the spectrum of NGC 1068:
   [Fe~X] $\lambda$6374 (Koski 1976), 
   [Fe~XI] $\lambda$7892 (Penston et al. 1984; Osterbrock \& Fulbright
   1996), and [Si~X] 1.430 $\mu$m (Thompson 1996).
   It has been recently suggested (Reynolds et al. 1997) that the coronal-line 
   emission arises in the outer regions of the X-ray absorber present
   in many Seyfert galaxies (see Reynolds (1997) and George et al. (1998)
   for a discussion of the properties of X-ray absorbers). 
   Furthermore, Krolik \& Kriss (1995) have postulated that the scattering
   medium in Seyfert 2 galaxies may be the same gas that
   produces the X-ray absorption in Seyfert 1s. Providing better constraints
   on the physical conditions in the gas in which the coronal lines arise will
   help in understanding the connection between the emission-line gas 
   and the scattering medium and, possibly, the X-ray absorber.   

   In this paper we will examine the physical conditions near Cloud B and
   the optical hot spot in the nucleus of NGC 1068. Among our
   results, we report the  
   presence of the highest ionization energy UV/optical lines ever 
   detected in a Seyfert 2 galaxy. In Section 3 we will discuss the observed relationship between ionization
   potential and the velocity of the emission-lines with respect to the
   systemic velocity of the host galaxy. In Sections 4 and 5 we will
   present the details of photoionization models of the emission-line
   gas. Finally we will discuss the relation between the emission-line
   gas, the scattering of the continuum radiation, and the structure
   of the inner NLR. 

\section{Observations and Analysis}

We obtained STIS long-slit spectra of NGC 1068 over 1150 -- 10,270 \AA~
on 1998 August 15. Paper I shows the slit position and describes the
observations and data reduction. The spectra that are analyzed in this paper
are from the central 0\arcsecpoint2 x 0\arcsecpoint1 bin in Paper I, which 
includes the brightest portion of the
hot spot in our slit and a portion of knot B south of its centroid.
We measured the fluxes of most of the narrow emission lines by direct 
integration over a local baseline determined by linear interpolation between 
adjacent continuum regions. For severely blended lines such as H$\alpha$ and [N~II] 
$\lambda\lambda$6548, 6584, we used the [O~III] $\lambda$5007 profile as a 
template to deblend the lines (see Crenshaw \& Peterson 1986).
We then determined the reddening of the narrow emission lines from the observed 
He~II $\lambda$1640/$\lambda$4686 ratio, the Galactic reddening curve of Savage 
\& Mathis (1979), and an intrinsic He II ratio of 7.2, which is expected from 
recombination (Seaton 1978) at the temperatures and densities typical of the NLR 
(see also Section 4).
We determined errors in the dereddened ratios from the sum in quadrature of the 
errors from three sources: photon noise, different reasonable continuum 
placements, and reddening. 

Table 1 gives the observed and dereddened narrow-line ratios, relative to 
H$\beta$, and errors in the dereddened ratios for each position. 
At the end of the table, we give the H$\beta$ flux
(ergs s$^{-1}$ cm$^{-2}$) in the bin and the reddening value that we 
determined from the He II ratios.

\section{Spectral Properties of the Gas Near the Hot Spot}

  Figure 1 shows the UV and optical spectra of the continuum hot spot
  in NGC 1068. Emission lines
  are present from a wide range in ionization state for the most numerous 
  elements, such as strong [N~II] $\lambda\lambda$6548, 6484 and
  N~V $\lambda$1240, and emission lines from the first four ionization
  states of oxygen, as seen in the earlier FOS spectra (Kraemer, Ruiz,
  \& Crenshaw; hereafter KRC).
  The continuum shows no strong evidence of a stellar component, and
  is clearly the result of scattered continuum radiation from the
  hidden central source, as we discussed in Paper I. Also,
  the strongest permitted lines show broad wings,
  which is due to reflected broad-line emission. 

  Netzer (1997) attributed the apparent weakness of O~III] $\lambda$1663
  in the FOS spectra of NGC 1068 to an underabundance of oxygen. On the
  other hand, we used the ratio of 
  N~III] $\lambda$1750/O~III] $\lambda$1663 as evidence of
  an overabundance of nitrogen (KRC). However,
  it is clear from these STIS data that O~III] $\lambda$1663 has been
  absorbed by Galactic Al~II $\lambda$1671 (see Figure 1) and, therefore,
  cannot be used to estimate either the O/H or N/O abundance ratios.

  The most intriguing feature of these spectra, which was not readily
  apparent in the FOS data, is the presence of a number of 
  coronal lines, including those from extremely high ionization states.
  As shown in Table 1, we have confirmed the presence of [Fe~X] $\lambda$6374
  and [Fe~XI] $\lambda$7892, which had been previously detected, as noted
  in Section 1. Furthermore, we have unambiguously detected 
  [Fe XIV] $\lambda$5303 (see Figure 1), which previously had only been 
  confirmed
  to be present in the spectra of the Seyfert 1 galaxies III Zw 77 (Osterbrock
  1981) and MCG -6-30-15 (Reynolds et al. 1997) and, possibly, the
  Seyfert 2 galaxy Tololo 0109 -383 (Fosbury \& Samsom 1983;
  Durret \& Bergeron 1988). Perhaps more unexpected
  is the presence of [S~XII] $\lambda$7611 (see Figure 1). S$^{+10}$ has an
  ionization energy of 504.7 eV, which makes this highest ionization
  line ever detected in the spectrum of a Seyfert galaxy, outside the
  X-ray region, and certainly the highest ever seen in the NLR, besting 
  [Si X] 1.43 $\mu$m, which has an ionization energy of 401.4 eV
  (Thompson 1996). Since 7611 \AA~ is within one of the O$_{2}$ telluric
  bands, detection is difficult from the ground. In fact, the
  only {\it confirmed} detection of [S~XII] has been in observations
  of the solar corona during the total eclipse of 30 May 1965 (Jefferies,
  Orrall, \& Zirker 1971). Osterbrock (1981) detected a line at a measured wavelength
  of 7613.1 \AA~ in the spectrum of III Zw 77, but did not
  identify it; we would suggest that this is also [S~XII]. 
    
  In addition to the above-mentioned lines, we have identified a number
  of other coronal lines in these spectra, as indicated by the ionization
  potentials listed in
  Table 1 (for the purposes of this paper, we refer to 
  ions with ionization energies greater than 100 eV as ``coronal'';
  this leaves out the [Ne~V] and [Fe~VII] lines, which, although
  detected in the solar corona, are formed under more typical NLR conditions).
  To our knowledge, these other lines have not been previously identified in the
  spectra of Seyfert galaxies, although we would expect that they should be
  present, since their ionization energies are lower than that of Fe$^{+13}$.
  Exceptions, perhaps, are the tentatively identified nickel lines,
  [Ni~XV] $\lambda$6702 and [Ni~XIII] $\lambda$5116, since the other
  lines are from more abundant elements. However, Halpern \& Oke (1986)
  detected [Ni~II] $\lambda$7378 in an off-nuclear spectrum of NGC 1068, and
  determined that the nickle abundance may be at least 4 times solar.
  One of the interesting aspects of the coronal lines is that they are
  blueshifted with respect to the lower ionization lines, an effect
  first noted in a number of Seyfert galaxies by Penston et al. (1984).
  In Figure 2, we plot ionization
  potential versus recession velocity (c$z$) for all the observed emission lines.
  Note that the presence of a blueshifted absorption feature can 
  bias the centroiding of the line redward, as appears to be the case for
  N~V $\lambda$1240 and C~IV $\lambda$1550 (KRC). Similarly, 
  the velocity measurement is more difficult for severly blended lines.
  Nevertheless, there is a strong correlation between ionization potential
  and velocity.
  A simple explanation is that the lines originate in different
  regions that are superimposed, such that we are viewing components with 
  different velocities along our line-of-sight. Note that most of the
  lower ionization lines are redshifted with respect to the systemic
  velocity of the host galaxy. 

  Figure 3 shows spatial cross-cut profiles of the [Fe XIV] $\lambda$5303 
  and [S~XII] $\lambda$7611 lines, along the slit. Clearly, the coronal line emission is 
  concentrated directly on the optical hot spot, in contrast to the 
  suggestion that it extends throughout the NLR in Seyfert
  galaxies (Korista \& Ferland 1989), but in general agreement with 
  Penston et al. (1984) regarding the origin of the [Fe~X] $\lambda$6374 
  and [Fe~XI] $\lambda$7892 lines, and results from line profile studies (cf. Moore,
  Cohen, \& Marcy 1996). Furthermore, the coronal line emission is 
  clumpier and not as extended as the continuum radiation. Of particular
  interest is the region NE of the hot spot, where the
  scattered continuum is strongest (see Paper I). 
  This does not necessarily mean that the coronal lines cannot arise
  in the same gas that scatters the continuum radiation, but, if they do,
  conditions in the scatterer must vary such that the lines are often
  weak. The possible connection between the scatterer and the coronal gas
  is discussed in Section 7.1.

\section{Photoionization Models}

\subsection{Abundances and Ionizing Continuum}   

   Our approach in photoionization modeling of NGC 1068 was
   described in detail in KRC. We will not repeat the details in the 
   current paper, except to point out important differences.
   As usual, the models are parameterized in terms of the dimensionless
   ionization parameter, U, which is the number of ionizing photons
   per hydrogen atom at the illuminated face of the cloud.
   Since the lines of Ne$^{+4}$ and Fe$^{+6}$ are quite strong in 
   in the spectrum of NGC 1068, we had assumed
   that this is in part due to an overabundance of these elements 
   relative to solar (cf. Oliva 1997). We still consider this to be true for iron, since
   {\it ASCA} data indicate a large Fe/O ratio in the X-ray
   emitting gas (Netzer \& Turner 1997). However, there is reason to believe
   that the Ne/H ratio is not substantially greater than solar at least
   in the extreme inner NLR. The
   strength of the [Ne~V] $\lambda$3426 relative to hydrogen is
   sensitive to the optical thickness of the emission-line gas. That
   is, if a cloud does not extend much beyond the He$^{++}$ zone,
   [Ne~V] $\lambda$3426/H$\beta$ can be quite large ($\sim$ 5). Also,
   if the ionizing continuum is somewhat harder than we had assumed,
   there are more photons capable of ionizing Ne$^{+3}$.

   In KRC, we had made the argument for a supersolar abundance of
   nitrogen in the inner NLR of NGC 1068, based on the strength of 
   N~V $\lambda$1240 relative to He~II $\lambda$1640 and C~IV $\lambda$1550 (cf. Ferland
   et al. 1996). However, the models significantly
   overpredicted N~IV] $\lambda$1486, which might indicate that
   that we overestimated the nitrogen abundance.
   The relative strength of N~V $\lambda$1240, and other
   resonance lines, can be boosted
   somewhat by fluoresence and scattering of ultraviolet continuum
   radiation from the central source (cf. Grandi 1975a, b). 
   As we will discuss in Section 6.1, continuum scattering and
   fluorescence cannot fully explain the strength of N~V $\lambda$1240.
   However, based on the observed N~IV] $\lambda$1486/H$\beta$ ratio,
   there is no reason to believe that the relative abundance of nitrogen
   is significantly greater than solar.

   Thus, other than iron, we have chosen to assume solar 
   abundances (cf. Grevesse \& Anders 1989) for these models. The numerical abundances,
   relative to hydrogen, are as follows: He=0.1,
   C=3.4x10$^{-4}$, O=6.8x10$^{-4}$, N=1.2x10$^{-4}$, Ne=1.1x10$^{-4}$, 
   S=1.5x10$^{-5}$, Si=3.1x10$^{-5}$, Mg=3.3x10$^{-5}$, Fe=8.0x10$^{-5}$.

   In KRC, we assumed an ionizing 
   continuum similar to that
   proposed by Pier et al. (1994), which is a conservative
   power-law fit from the UV and to the X-ray, using the 2 keV flux
   and X-ray continuum derived from the BBXRT data (Marshall et al. 1993). 
   However, fits to the X-ray continuum by combining
   {\it ROSAT}/PSPC and {\it Ginga}, or {\it Einstein}/IPC
   and {\it EXOSAT} data (see Pier et al. 1994, and 
   references therein) are consistent with a somewhat harder spectral energy distribution (SED). 
   As noted above, the strength of [Ne~V] $\lambda$3426 depends on the
   hardness of the ionizing continuum. Therefore, we modeled
   the SED as a broken power-law of the form, F$_{\nu}$ $=$ K$\nu^{-\alpha}$, 
   as follows:

\begin{equation}
    \alpha = 1.0,~ h\nu < 13.6 eV
\end{equation}
\begin{equation}
    \alpha = 1.4,~ 13.6eV \leq h\nu < 1000eV
\end{equation}
\begin{equation}
    \alpha = 0.5, ~h\nu \geq 1000eV
\end{equation}
   Note that we have assumed a harder EUV - Xray continuum than 
   in KRC (1.4, rather than 1.6) and have
   positioned the X-ray break at one-half the energy.
   We have, however, assumed the same intrinsic luminosity above the
   Lyman limit, Q $=$ 4x10$^{54}$ photons 
   sec$^{-1}$ (Pier et al. 1994), which is typical of Seyfert 1 nuclei .

\subsection{Component Parameters}

   As noted in Section 3, these spectra show emission from a wide
   range of ionization states, which is an indication that we are seeing
   emission from a range of physical conditions. Also, there is an
   apparent correlation between the redshift of the emission lines
   and the ionzation potential of the ion from which they arise.
   Thus, it is likely that we are observing emission from several
   distinct components.

   An initial guess at temperature and density of much of the
   emission-line gas (which we will refer to as the HIGHION component) can be derived from the 
   ratio of [O~III] $\lambda\lambda$5007,4959/[O~III] $\lambda$4363 
(Osterbrock 1989). The dereddened observed ratio is $\sim$ 47, indicating 
   an electron temperature T$_{e}$ $\leq$ 20,000K in the low density limit, and little
   modification by collisional de-excitation of the $^{1}$D$_{2}$ level.
   [Fe~VII] lines ratios can also be used to estimate temperature.
   The ratio [Fe~VII] $\lambda$3759/[Fe~VII] $\lambda$6087
   $\approx$ 0.79, which indicates T$_{e}$ $\sim$ 25,000K, 
   at electron densities, n$_{e}$, $\leq$ 10$^{6}$ cm$^{-3}$ (Nussbaumer \&
   Storey 1982). Still, it is possible
   that much of the [Fe~VII] and [O~III] emission arises
   in different zones within the same gas, particularly since these lines are at the same approximate
   velocity (see Table 1).  

   If one kinematic component contributes most of the [O~III] and
   [Fe~VII] emission, and presumably other lines within the same 
   range of ionization energy (see Table 1), it is not likely to contribute the
   highest ionization lines in these spectra. As we discussed in Section
   3, such lines as [Fe~IX] $\lambda$7892, [Fe~XIV] $\lambda$5303, and 
   [S~XII] $\lambda$7611, are blueshifted with respect to the lower ionization
   lines by several hundred km sec$^{-1}$, which suggests that they arise
   in a different kinematic component (which we will refer to as CORONAL). It has been previously
   suggested that a rough 
   estimate of the temperature of the coronal gas can be obtained from the
   ratio of [Fe~XI] $\lambda$2649/[Fe~XI] $\lambda$7892
   (Penston et al. 1984; Osterbrock \& Fulbright 1996). The observed ratio is
   [Fe~XI] $\lambda$2649/[Fe~XI] $\lambda$7892 $=$ 1.4 $^{+.43}_{-.35}$. Unfortunately,
   based on the most recent calculations for effective collision strengths 
   (Tayal
   1999), this ratio is not particularly sensitive to temperature
   at electron densities $\leq$ 10$^{8}$ cm$^{-3}$. Using these collision strengths
   and the transition probabilities from Mason (1975), the 
   dereddened [Fe XI] ratio 
   indicates that T$_{e}$ $>$ 3 x 10$^{4}$ K in the Fe$^{+10}$ zone.
   Since the 
   redshifts of the
   Balmer lines are similar to those of the [O~III] lines, we expect
   CORONAL will contribute little to the total hydrogen recombination line
   flux.

   Since the lowest ionization lines appear redshifted compared to the 
   other emission-lines, it is probable that they arise in a separate
   component, rather than in the more neutral parts of HIGHION.
   This
   component must be of fairly low density, since the ratio
   of [S~II] $\lambda$6716/[S~II] $\lambda$6731 $\approx$ 0.81,
   indicating n$_{e}$ $\sim$ 2 x 10$^{3}$ cm$^{-2}$ (Osterbrock 1989). 
   The S$^{+}$ lines can arise in partially neutral gas and, thus, we estimate
   that the atomic hydrogen density for this component is somewhat
   higher, n$_{H}$ $\leq$ a few times 10$^{4}$ cm$^{-2}$. The nature of
   such a component 
   (which we will refer to as LOWION) was discussed extensively in KRC. 
   The fact that
   the strengths of the low-ionization, collisionally-excited lines must be large relative
   to the Balmer lines in order for them to appear strong in 
   a composite spectrum led us to believe that the gas was screened
   and, thus, ionized by an absorbed continuum strongly weighted to the X-ray
   (KRC). However, the paucity of ionizing photons in the absorbed
   continuum and the constraint that this component had the same
   covering factor as the screening gas led to an
   underprediction of its contribution to the total emission-line flux.
   In these spectra it is apparent that much of the high ionization
   gas is optically
   thin, since the observed He~II $\lambda$4686/H$\beta$ ratio is 0.60, which
   is unattainable if much of the gas is optically thick near the Lyman limit.
   Thus, neither of the other components can
   provide an effective screen for LOWION. 

   Kraemer et al. (1999a) have modeled the effect on the NLR of an absorber 
   with a high covering factor 
   within a few parsecs of the continuum source, and this effect is clearly evident 
   in the NLR spectrum of NGC 4151 (Alexander et al. 1999; Kraemer et al
   1999b). Thus, we suggest the following geometry for the inner
   NLR of NGC 1068. The components represented by HIGHION and CORONAL are essentially
   co-located, and are ionized by unabsorbed continuum radiation
   from the hidden nucleus. The low ionization gas lies out of the plane
   occupied by HIGHION and CORONAL, and is irradiated by X-rays
   which penetrate a layer of absorbing gas closer to the nucleus.
   Also a small fraction of 
   unabsorbed continuum, scattered
   out of the plane by free electrons, is incident upon the low ionization gas.
   This simple geometry is illustrated in 
   Figure 4. There is no simple way to quantify the fractions of scattered and direct
   continuum irradiating LOWION, but it is thought that at least
   1\% of the continuum radiation in NGC 1068 is scattered into our
   line of sight (Miller et al. 1991) and there is the
   example of a large column, X-ray absorber in NGC 4151 (cf. Barr et al. 1977) 
   which may vary in optical depth as a function of angle with respect
   to the ionization cone (Kraemer et al. 1999b). 
   We have modeled
   the absorber assuming an ionization parameter, U $=$ 0.1, and a column density, 
   N$_{H}$ $=$ 7.4 x 10$^{22}$ cm$^{-2}$, where N$_{H}$ is the sum
   of the columns of ionized and neutral hydrogen. These parameters
   are similar to those describing the X-ray absorber in NGC 4151
   (Yaqoob, Warwick \& Pounds 1989; Weaver et al. 1994). Given our
   estimate of the luminosity of the active nucleus, 
   an absorber of density  n$_{H}$ $=$ 10$^{7}$ cm$^{-3}$ would 
   lie at a distance of $\sim$ 1 pc from the central engine, 
   closer to the nucleus than the inner part of hot spot.
   The ionizing continuum incident upon LOWION is shown in Figure 5.
    For the purposes of these 
   models, we assume that HIGHION and LOWION are at the same radial distance 
   from the nucleus. 
  
   Based upon the observed physical properties derived from the
   emission-lines and the
   simple geometric picture discussed above, we have generated 
   a three-component photoionization model to describe the physical    
   conditions of line emitting gas. Most of the emission arises
   in the component HIGHION, with n$_{H}$ $=$
   6 x 10$^{4}$ cm$^{-2}$, U $=$ 10$^{-1.52}$, and ,
   N$_{H}$ $=$ 10$^{21}$ cm$^{-2}$. The ionization parameter was chosen
   to produce strong [Fe~VII] emission with negligible
   [Fe~XI] $\lambda$7892. The model was truncated at the 
   termination of the He$^{+2}$ zone, in order to produce the
   large He~II $\lambda$4686/H$\beta$ ratio and strong [Ne~V] described above. 
   The density was constrained on the high end by the presence of strong 
   [Ne~IV] $\lambda$2423, which has a critical 
   density $\sim$ 10$^{5}$ cm$^{-3}$,
   and on the low end by the ionization parameter.
   LOWION is characterized by n$_{H}$ $=$ 3 x 10$^{4}$ cm$^{-2}$, U $=$ 10$^{-3.2}$
   (from the combined continuum). Since the size of the partially neutral 
   zone in X-ray ionized gas can be inordinately large, we
   chose to truncate the model at the same column density as HIGHION. 
   Although not completely radiation bounded, LOWION has a considerable 
   extended zone behind the H$^{+}$/H$^{0}$ boundary, from which 
   [O~I] $\lambda\lambda$6300, 6364, Mg~II $\lambda$2800, 
   and [S~II] $\lambda\lambda$6716, 6731 arise. For LOWION, we
   used a solar iron abundance, since the enhanced cooling by [Fe~II] in 
   the extended zone would suppress other lines. The lack of reliable
   atomic constants for coronal gas at nebular temperatures (i.e., T$_{e}$ $\leq$
   10$^{5}$K) makes it difficult to set the model input parameters for
   CORONAL on the basis of emission-line ratios. We
   have generated this component using CLOUDY90 (Ferland et al 1998)
   since it includes more a complete model for the coronal-line
   emission, and includes elements, specifically argon and nickel,
   which are not included in our code. Therefore, we
   generated a single coronal component to produce the mix of ionization
   states observed in the blue-shifted emission-lines, assuming the
   following: n$_{H}$ $=$ 7 x 10$^{2}$ cm$^{-3}$, U = 1.7, N$_{H}$ = 
   4 x 10$^{22}$ cm$^{-2}$.

   In Table 1, the dereddened L$\alpha$/H$\beta$ is 30.75, which is roughly
   equal to the value from recombination plus collisional excitation
   in low ionization gas (Osterbrock 1989). Thus, there is no
   evidence for the destruction of trapped L$\alpha$ photons
   by dust. Furthermore, dust is not responsible for 
   scattering of continuum radiation by the hot spot (see Paper I), 
   which is another indication of the absence of dust in this region. 
   Therefore, we have assumed that the emission-line gas is dust-free.

\section{Model Results}

   In creating a composite model, we first scaled the contributions
   from HIGHION to provide a rough fit to the high ionization
   lines such as C~IV $\lambda$1550, [Ne~V] $\lambda$3426, and
   [Fe~VII] $\lambda$6087 and LOWION to fit lines such as [N~II] $\lambda$6584, 
   [O~II] $\lambda$3727, with the result that the contribution of HIGHION 
   to the total H$\beta$ flux is 3 times that of LOWION. Due to the uncertainties in the
   atomic data, we have elected not to include
   the predicted forbidden line strengths from CORONAL in our scaling.
   Since the Balmer lines are associated kinematically with the
   non-coronal emission lines, it seems reasonable to expect
   that the contribution from CORONAL is $\leq$ 15\% of the total H$\beta$ flux,
   which is only slightly more than the uncertainty in measurement.

   In Table 2, we compare the predicted line ratios for HIGHION and LOWION 
   and the composite model to the observed/dereddened values.
   Given the simplicity of the model, we have obtained very
   satisfactory fits for the vast majority of the observed
   emission lines. There is good agreement over
   nearly the full ionization sequence, for example C~IV $\lambda$1550, 
   C~III] $\lambda$1909, and C~II] $\lambda$2325, and O~IV] $\lambda$1402,
   [O~III] $\lambda$5007, [O~II] $\lambda$3727, and [O~I] $\lambda$6300.
   The predicted [O~III] $\lambda\lambda$5007,4959/[O~III] $\lambda$4363 
   ratio is 46, which indicates that T$_{e}$ in the O$^{+2}$, averaged
   over the two components, is correct. The [Fe~VII] $\lambda$3759/[Fe~VII] 
   $\lambda$6087 ratio is 0.55, lower than the observed value,
   which indicates a somewhat higher temperature in the 
   Fe$^{+6}$ zone than predicted, which is not surprising since
   given our model requirement that this zone is co-located with the
   O$^{+2}$ zone. The predicted [S~II] $\lambda$6716/$\lambda$6731
   ratio is 1.15, identical to that observed within the errors,
   which confirms our assumptions regarding the density and ionization
   structure of LOWION. In general, the model predictions demonstrate
   that our assumptions regarding the elemental abundances are
   approximately correct. However, the models do underpredict the strengths
   of the neon lines somewhat, which may indicate that the neon is 
   supersolar, but probably less than a factor of two. 

   The model prediction for the He~II $\lambda$4686/H$\beta$ and ratio
   is only slightly higher than observed. The predicted strengths of the lines
   formed in the partially neutral envelope of the low ionization
   gas, such as [S~II] $\lambda\lambda$6716, 6731 and [O~I] $\lambda$6300,
   are in reasonable agreement with the observations.
   This indicates that the combined
   effects of SED and column density are well represented by the models.
   This result is of particular importance, given the assumption that
   the components are ionized by different continua. 

   Also listed in Table 2 are the model predictions for 
   the emitted H$\beta$ flux, the emitting surface area (the 
   scaled to the reddening-corrected H$\beta$ luminosity divided by the 
   emitted flux), and
   covering factor for each component, assuming a distance of $\sim$
   25 pc from the hidden continuum source. At the distance of
   NGC 1068 (14.4 Mpc, Bland-Hawthorne 1997), the 0\arcsecpoint1 slit 
   width corresponds to 7.2 pc, yielding a covering factor for the slit of
   $\sim$ 0.05. Since the covering factor of these components are
   substantially lower than 0.05, there is no evidence that we
   are seeing substantial effects of superimposition of clouds along our
   line-of-sight.
   
   Our photoionizaton code does not include pumping of UV resonance
   lines by scattering of continuum radiation and continuum fluoresence
   (cf. Ferguson, Ferland,
   \& Pradhan 1994), which may explain the underpredictions of the strengths of several UV 
   resonance lines, including N~V $\lambda$1240, C~II $\lambda$1335 and Mg~II
   $\lambda$2800.
   Therefore, we recomputed the LOWION model using CLOUDY90, 
   assuming a turbulent velocity of 50 km s$^{-1}$. 
   Although pumping of UV resonance lines is most efficient for gas with
   large turbulent velocities ($\geq$ 1000 km s$^{-1}$; Ferguson et al. 1994),
   it can still be an important process 
   if the covering factor of the emitting gas is sufficiently large 
   (cf. Hamann \&
   Korista 1996) and/or if the optical depths of the scattered lines are
   small (Ferland 1992). We derive a relatively large
   covering factor for LOWION ($\sim$ 20\% that of the slit), and thus it is not 
   surprising that the model predicts a significant contribution
   to resonance lines from scattered continuum radiation (interestingly,
   this is all due to 
   direct pumping of
   the UV resonance line, since there is insufficient ionizing radiation
   incident upon LOWION to pump the EUV driver lines).
   The CLOUDY90
   predictions for C~II $\lambda$1335 and Mg~II $\lambda$2800 are listed
   in Table 2, alongside those from our code (the predictions for
   the non-resonance lines were quite similar for the two codes), and
   the agreement with observed flux ratios is quite good.
   In addition to the large column densities of
   C$^{+}$ (3.22 x 10$^{17}$ cm$^{-2}$) and Mg$^{+}$ 
   (1.54 x 10$^{16}$~cm$^{-2}$), LOWION predicts a large column density
   for O$^{0}$ (6.02 x 10$^{17}$ cm$^{-2}$). O~I $\lambda$1302 is present in 
   the far-UV 
   spectrum (see Figure 1) and we expect that is also formed by 
   continuum scattering (unfortunately, it is not included in the
   code output). However, continuum scattering does not appear to have a 
   similarly strong affect in boosting the N~V $\lambda$1240 line, as
   we discuss below.
   
   In Table 3, we list the predictions of the CORONAL model for the mean 
   ionization fractions; ions with observed lines are flagged.
   We elected not to list the predicted emission-line
   ratios since they may be misleading given the lack of 
   reliable atomic data (cf. Moorwood et al. 1997). The model predicts 
   non-negligible 
   populations for each of the observed ionic states, except Si$^{+6}$ and
   Fe$^{+6}$, however it is clear that the [Fe~VII] lines arise in lower
   ionization gas.
   If we assume that CORONAL contributes 15\% of the observed H$\beta$
   flux, the covering factor is 2.2 x 10$^{-3}$. Assuming isotropic
   scattering, at small electron scattering optical
   depths, $\tau$$_{electron}$ $<$ 1, the reflected
   fraction of continuum radiation, f$_{refl}$ $\approx$ 
   N$_{e}$F$_{c}$$\sigma_{Thomson}$,
   where N$_{e}$ is the column density of electrons, F$_{c}$ is the covering
   factor, and $\sigma_{Thomson}$ is the Thomson cross-section. For CORONAL,
   f$_{refl}$ $\approx$ 7.4 x 10$^{-5}$. However,  
   the observed reflected continuum fraction in our spectrum, based on the estimated
   central source luminosity (Pier et al. 1994), is
   f$_{refl}$ $\sim$ 1.4 x 10$^{-3}$
   (the total reflected fraction is  f$_{refl}$ $\approx$ 1.5 x 10$^{-2}$,
   consistent with the larger region sampled; Miller
   et al. 1991). Thus, the coronal-line emitting
   gas near the hot spot makes a neglible contribution
   to the scattered continuum radiation.

\section{Open Issues}

\subsection{The Nitrogen Problem}

   In KRC, we attributed the unusual strength of the N~V $\lambda$1240
   line to an overabundance of nitrogen, however, the models significantly
   overpredicted N~IV] $\lambda$1486. Even with solar nitrogen abundance,
   HIGHION overpredicts N~IV] $\lambda$1486 by a factor of nearly 3.
   A check of HIGHION using CLOUDY90, assuming a turbulent velocity 
   of 500 km s$^{-1}$, produced only a 70\% increase in the
   strength of N~V $\lambda$1240. The 
   models do not significantly underpredict other lines formed in the same 
   zone as N~V $\lambda$1240, such as [Ne~V] $\lambda$3426 and 
   [Fe~VII] $\lambda$6087, and the predictions for other UV resonance lines
   match the data or are well understood (see above). One possible 
   explanation is that N~V $\lambda$1240 and N~IV] $\lambda$1486 are formed 
   in different regions, with much different nitrogen abundances,
   but we find no evidence for abundance inhomogeneities for
   elements other than iron. Therefore, the problem with the
   nitrogen lines remains unresolved.

\subsection{Coronal Component}
   
   The predicted temperatures for CORONAL 
   (T$_{e}$ $=$ 9.29 x 10$^{4}$ K at the ionized face, 3.67 x 10$^{4}$ K at
   the point of truncation),
   are within the range expected from the 
   [Fe~XI] $\lambda$2649/$\lambda$7892
   ratio. However, since the [Fe~XI] ratio is not 
   particularly sensitive to temperature at low density, the 
   estimate of the coronal gas temperature is not well constrained.
   Hopefully, better estimates may be obtained once more accurate 
   collision
   strengths are calculated for other coronal lines in the temperature range 
   10$^{4}$ -- 10$^{5}$ K.
   In any case, it is clear from our model predictions that the
   observed ionic states can co-exist in photoionized gas 
   characterized by a single ionization parameter.

\section{Discussion}

\subsection{Coronal Gas, the Scattering Medium, and X-ray Absorbers}

   As we have demonstrated, the ionic states indicated by the
   coronal emission-lines can co-exist in a large column of highly
   ionized gas. The ionization parameter, U $=$ 1.7, is characteristic
   of the X-ray absorbers present in many Seyfert 1 galaxies (cf.
   Reynolds 1997). It is tempting to associate the coronal-line
   emitting gas with the X-ray absorber, as Reynolds et al. (1997) has
   done for the case of the Seyfert 1 galaxy, MCG -6-30-15. Based on the
   fraction of Seyfert 1s that possess an X-ray absorber, 
   it is likely that the covering factor of the absorber
   is 0.5 -- 1 (cf. George et al. 1998). However, we find that the covering 
   factor of the coronal-line gas in NGC 1068 is quite low, and thus
   different from typical X-ray absorbers.

   The low covering factor of the coronal gas also makes it 
   unlikely to be the continuum scattering region. Furthermore,
   Miller et al. (1991) estimated a temperature for the scattering
   medium, T$_{e}$ $\sim$ 3 x 10$^{5}$ K, which is significantly greater
   than our model predictions. Therefore, it is likely that the scattering
   occurs in a component of 
   gas associated with the hot spot that is more highly
   ionized than our coronal component. If we use the initial
   conditions, n$_{H}$ $=$ 200 cm$^{-3}$, U $=$ 8, 
   and N$_{H}$ $\sim$ 10$^{22}$ cm$^{-2}$, CLOUDY90 predicts
   a mean T$_{e}$ $\sim$ 4.5 x 10$^{5}$~K, which is close to
   Miller et al.'s value. If we, again, apply the constraint that this
   additional component contributes $\leq$ 15\% of the total H$\beta$,
   the covering factor for this component is $\leq$ 0.07 and, thus,
   f$_{refl}$ $\leq$ 1.2 x 10$^{-3}$, or approximately 85\% that
   observed. We would expect to see some line emission from this component,
   but only from the most highly ionized species, such as [S~XII] $\lambda$7611.
   The covering factor for this component is slightly greater than that 
   constrained by the slit width, which would not
   be surprising if the scatterer is indeed an X-ray absorber viewed across our 
   line-of-sight, as suggested by Krolik \& Kriss (1995), and there was
   some superposition of clouds.

   To summarize, we think it is unlikely that the coronal-line
   gas has a sufficient covering factor to produce the scattered
   continuum radiation. It is plausible that the scattering occurs
   in a component of more highly ionized gas, with a high covering
   factor, which may contribute a fraction of the coronal-line 
   emission. Although the physical conditions of both components 
   are within the range observed for X-ray absorbers (cf. Reynolds 1997),
   we suggest that while the absorber may be associated with the scatterer, 
   neither are associated with the coronal-line gas that we observe.
   
\subsection{Photoionization versus Collisional Processes}

   The relative contributions of photoionization and collisional
   processes (e.g., shocks, heating by cosmic rays) to the physical state of 
   the emission-line gas in NGC 1068 has been a matter of some debate.
   While Kriss et al. (1992) have attributed the strong C~III $\lambda$977
   and N~III $\lambda$990 seen in {\it HUT} spectra to shock
   heating, Ferguson et al. (1994) suggest that the strength of these
   lines result from a combination of continuum fluoresence and
   dielectronic recombination. 
   Given the importance of 
   continuum scattering to O~I $\lambda$1302, C~II $\lambda$1335 
   and Mg~II $\lambda$2800,
   it is not surprising that the same process enhances C~III $\lambda$977
   and N~III $\lambda$990. In fact, the recomputed LOWION predicts
   relatively strong C~III $\lambda$977 and N~III $\lambda$990 (see Table 2).
   The contribution to both lines from continuum scattering is
   $\sim$ 90\% for LOWION and $\sim$ 50\% in HIGHION, primarily from zones 
   near the ionized face
   of the cloud, where the driver lines are optically thin and
   pumping is most efficient (Ferland 1992). The models
   predict C~III] $\lambda$1909/C~III $\lambda$977 $\approx$ 5.2
   and N~III] $\lambda$1750/N~III $\lambda$990 $\approx$ 1.1, compared
   to 3.15 $\pm$ 0.51 and 1.46 $\pm$ 0.34, respectively, from the {\it HUT} spectra 
   (Kriss et al. 1992), noting that these ratios are 
   quite sensitive to the atomic parameters used in the code. Thus, although 
   we cannot rule out additional
   heating mechanisms, it is clear that continuum pumping can dramatically
   enhance these lines in photoionized gas. 
   
   Nevertheless, it is interesting to note that turbulent velocities as low
   as 50 km s$^{-1}$ can fully account for the resonance scattering.
   The full width at half maximum of C~II $\lambda$1335 is $\sim$ 1240 
   km s$^{-1}$,
   corrected for Galactic absorption, which indicates that the
   line is broadened by the summation of different kinematic 
   components. Interestingly, the widths of the resolved kinematic components
   of instrinsic UV absorbers in Seyfert 1 galaxies are also
   typically $\sim$ 50 km s$^{-1}$ (Crenshaw et al. 1999). The
   lack of large scale turbulence implies a lack of violent disruption of the
   gas.

   In KRC, we attributed the large H$\alpha$/H$\beta$ ratio
   in at least one region to collisional excitation of H$\alpha$
   by an injection of energetic particles, possibly associated with the radio
   jet, into the NLR gas. This effect was not as apparent 
   in the optical nucleus, and we find no evidence in
   our STIS data for enhancement of H$\alpha$ beyond the
   predictions of the photoinization models. However, there may
   still be examples of jet/cloud interaction at other locations
   in the inner NLR OF NGC 1068. This will be addressed in a 
   subsequent paper.

   One piece of evidence that heating processes other than photoionization
   are present is that the electron temperatures predicted by the
   models are somewhat lower than those estimated from the
   [Fe~VII]. However, the underprediction of the
   [Fe~VII] ratio is probably due to our assumption that these 
   lines arise in the same gas as the [O~III] lines.
   Therefore, while it is possible that
   collisional effects are important in some of the high ionization
   gas in the inner NLR, most of the observed properties
   are consistent with photoionization by the central source.

\subsection{Geometry of the Inner NLR}

   While there is no additional evidence to support our proposed 
   geometry for the inner NLR in NGC 1068, assuming that
   the low ionization gas is screened by a large column absorber
   resolves a problem with
   the KRC model regarding the covering factor of the screened
   gas. Also, it is apparent that conditions in the NLR of NGC 4151
   are due to absorption of the ionizing continuum by intervening
   gas (Alexander et al. 1999; Kraemer et al. 1999b). If a column
   of extremely optically thick gas is present, it may have an important
   effect in the collimation of the ionizing radiation.
   However, in order to do so, the absorber must have 
   a large covering factor (i.e., 0.5). 
   In addition to NGC 4151, a large column of 
   X-ray absorbing
   gas has been detected in the Seyfert 1.5, Mrk 6 (Feldmeier et al. 1999),
   but there are not enough examples to make a statistical
   determination of the covering factor of the
   absorber, or its column density
   as a function of scale-height.
   Nevertheless, if this component has a large
   covering factor, it will be important, along
   with the putative molecular torus and any intrinsic 
   anisotropy of the radiation field, in determining the 
   distribution of ionized gas in the NLR.

\section{Conclusions}

   We have analyzed the STIS UV and optical spectra of the inner nuclear
   region of NGC 1068, near the optical continuum peak or hot spot. 
   We have contructed photoionization models of the narrow-line
   gas and have been able to successfully match nearly all the
   observed dereddened emission-line ratios. From our analysis and
   modeling of these spectra, we can report the following
   discoveries regarding the physical conditions near the hot spot.

   1. We report the detection of a number of strong coronal emission
   lines, including [Fe~XIV] $\lambda$5303 and [S~XII] $\lambda$7611.
   The latter is the highest ionization UV/optical line ever identified in the
   NLR of a Seyfert galaxy. The coronal lines are blueshifted with
   respect to both the systemic velocity of the host galaxy and
   the lower ionization lines in these spectra. This indicates that
   the coronal lines arise in a distinct component of narrow-line
   gas. The lower velocities of other emission-lines
   may be due to contributions from gas at larger radial
   distances along our line-of-sight, but it is also possible
   that much of this gas is co-located with the coronal
   component, which implies that the velocity of the gas
   is related to its ionization state and/or density. If this 
   gas is in radial outflow from the nucleus, these observations
   may provide important constraints for models of cloud acceleration
   (the kinematics of the inner NLR of NGC 1068 will be the subject of
   a later paper).

   2. We have used a three-component model to match the narrow emission-line
   flux ratios, since the kinematics and range of physical conditions
   clearly indicate that we are sampling distinct emission-line clouds.
   Most of the high ionization lines, such as C~IV $\lambda$1550, 
   [Ne~V] $\lambda$3426 and [Fe~VII] $\lambda$6087, 
   arise in a component, HIGHION, which is directly ionized
   by the hidden continuum source and is optically thin at the Lyman limit.
   Low ionization lines, such as Mg~II $\lambda$2800, [O~II] $\lambda$3727,
   and [N~II] $\lambda$6584, are formed in a component, LOWION, which is 
   screened from the central source by an optically thick layer of gas (N$_{H}$
   $=$ 7.4 x 10$^{22}$ cm$^{-2}$), similar to that seen in NGC 4151.

   Several of the UV resonance lines formed in the low ionization
   gas (O~I $\lambda$1302, C~II $\lambda$1335, and Mg~II $\lambda$2800) are 
   enhanced by the scattering of continuum radiation. The turbulent
   velocities required to produce this effect are low, $\sim$ 50 km s$^{-1}$.
   Although this is greater than the thermal velocities within
   the emitting clouds, it is much less than one might expect if
   the clouds were being disrupted (e.g., by fast shocks). Also, it
   is an interesting coincidence that the required turbulent
   velocities are similar to those measured in the resolved 
   kinematic components of the intrinsic UV absorbers in Seyfert 1s.

   Unlike our previous study (KRC), we find no strong evidence
   for a large overabundance of neon and nitrogen, although
   the Fe/H ratio in the higher ionization gas appears to
   be approximately twice solar. Interestingly, there cannot be
   significant iron enhancement in the low ionization gas, otherwise
   the model underpredicts the electron temperature in the partially 
   neutral zone. Although iron could be depleted onto dust
   grains in the neutral gas, we find no evidence of the 
   effects of dust mixed in with the emission-line gas.
   
   Assuming solar abundances, the models underpredict 
   N~V $\lambda$1240 and overpredict N~IV] $\lambda$1486.
   Resonance scattering and continuum fluoresence cannot
   sufficiently pump the N~V line to match the observed flux,
   and, in any case, these processes do not affect the
   N~IV] strength. Although extremely inhomogeneous nitrogen
   abundances could cause this effect, we think that this
   is quite unlikely. 
   
   3. Although the atomic data is not sufficiently reliable to 
   predict coronal-line flux ratios, we have generated a component,
   CORONAL, in which all the observed ions are present. Hence, it
   is plausible that these lines form in photoionized gas. We do
   not believe that this component is responsible for the
   scattered continuum radiation, since its covering factor and 
   electron temperature are too low. As such, we postulate that a 
   more highly ionized, co-located component is the scatterer. This
   component has a covering factor and ionization state similar
   to the X-ray absorbers detected in Seyfert 1 galaxies.

\acknowledgments

  S.B.K would like to thank Swaraj Tayal, Don Osterbrock, Anand Bhatia, Dick Fisher, and 
  Bruce Woodgate for illuminating conversations about coronal emission-lines. 
  S.B.K.
  would also like to thank Gary Ferland for useful discussions about
  resonance scattering and continuum pumping of emission lines. We also
  thank Cherie Miskey for help with the figures.
  S.B.K. and D.M.C. acknowledge support from NASA grant NAG 5-4103.

\clearpage

\figcaption[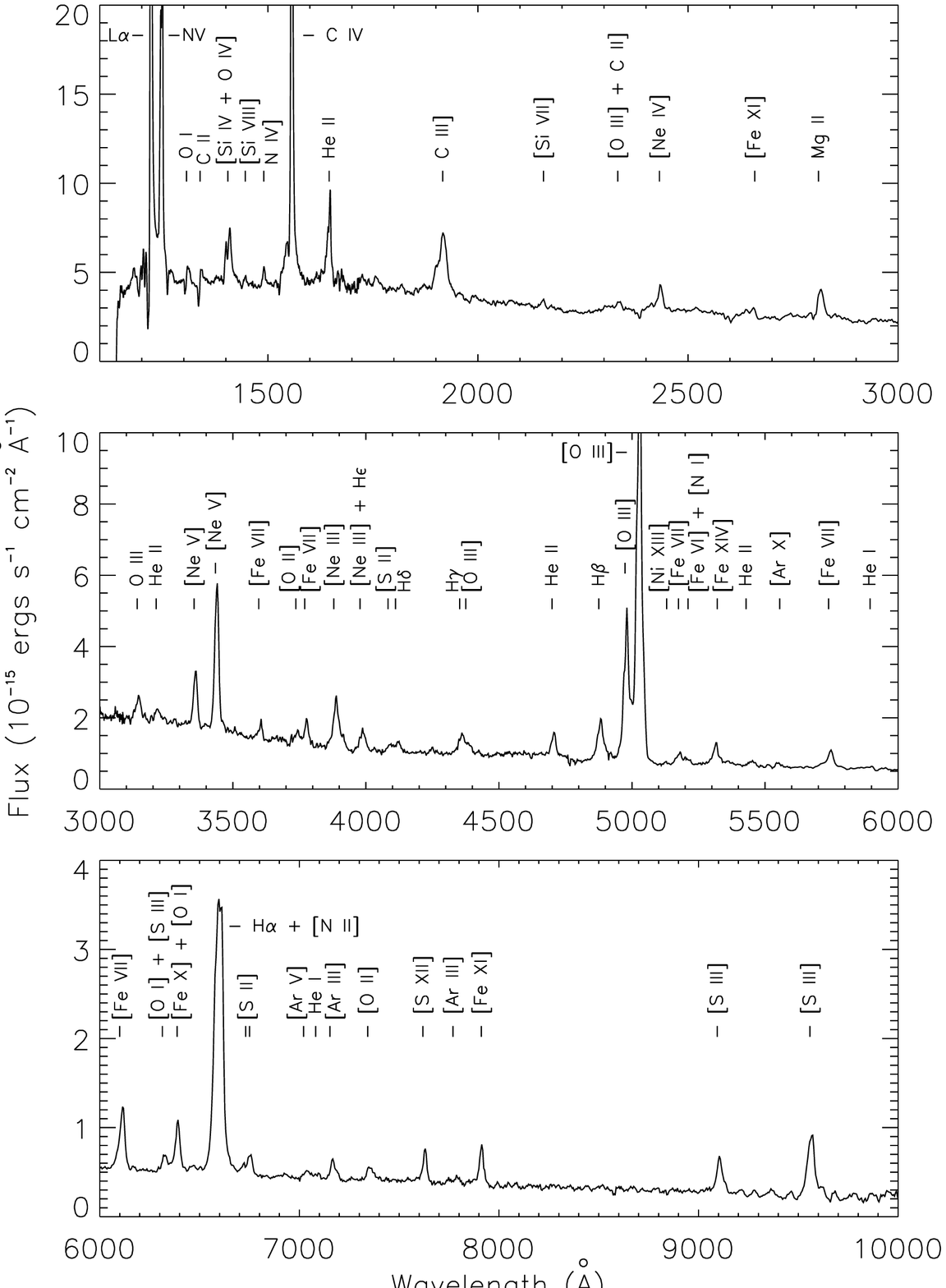]{STIS Far-UV (G140L) and Near-UV (G320L), top,
Blue Optical (G430L), middle, and Optical/Near-IR (G750L), bottom,
spectra of NGC 1068 near the nuclear hot spot.
}\label{fig1} 

\figcaption[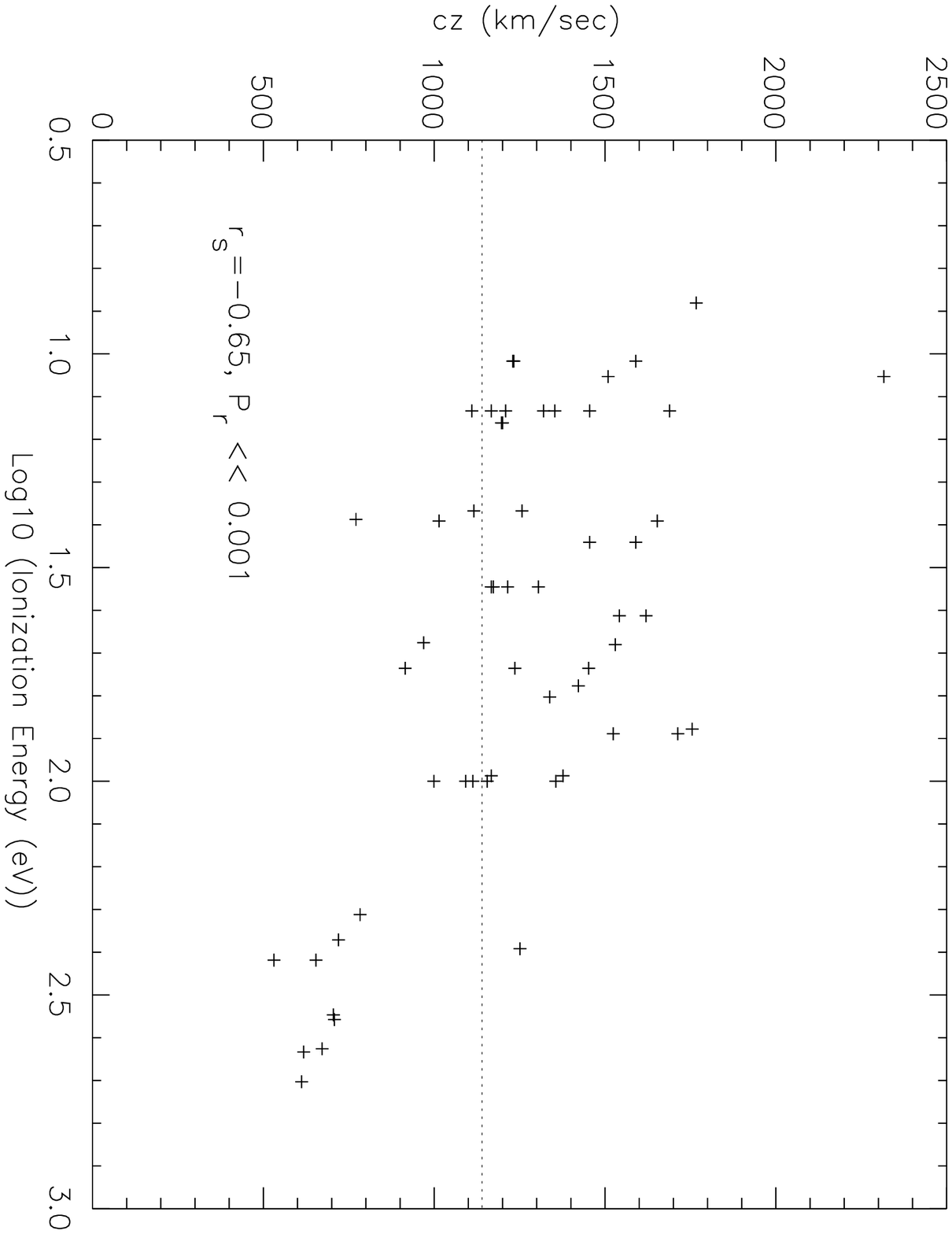]{Plot of the recession velocity (cz)
against the log of the ionization energy for the emission-line 
detected in the STIS spectra of the region near the hot spot.
As noted in the text, heavily absorbed and severely
blended lines have been omitted. The dotted line shows the systemic velocity.
}\label{fig2} 

\figcaption[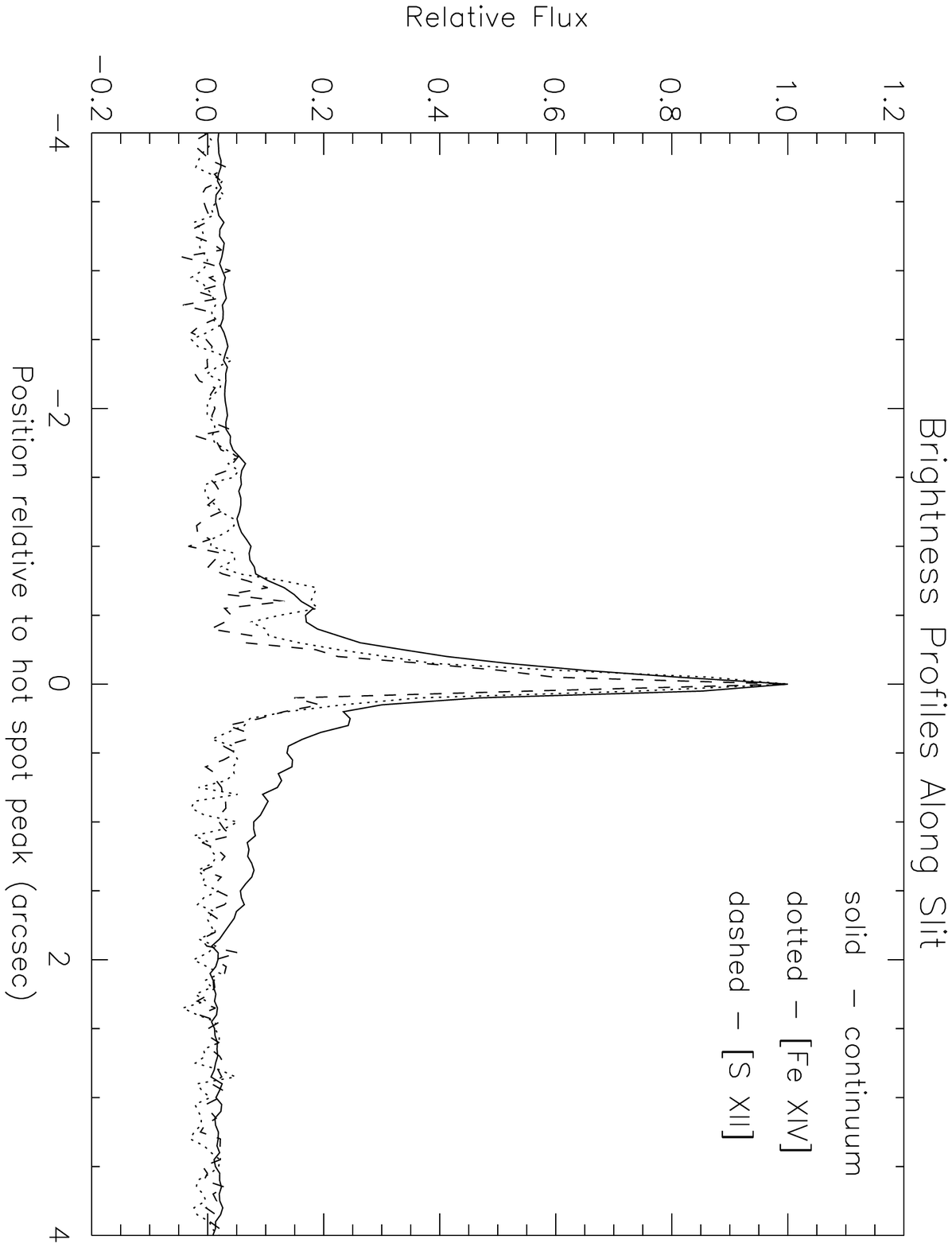]{Relative brightness profiles of the continuum
and [Fe~XIV] and [S~XII] emission along the slit. Negative positions
correspond to the NE direction (see Paper I for the details of the
observations).
}\label{fig3} 

\figcaption[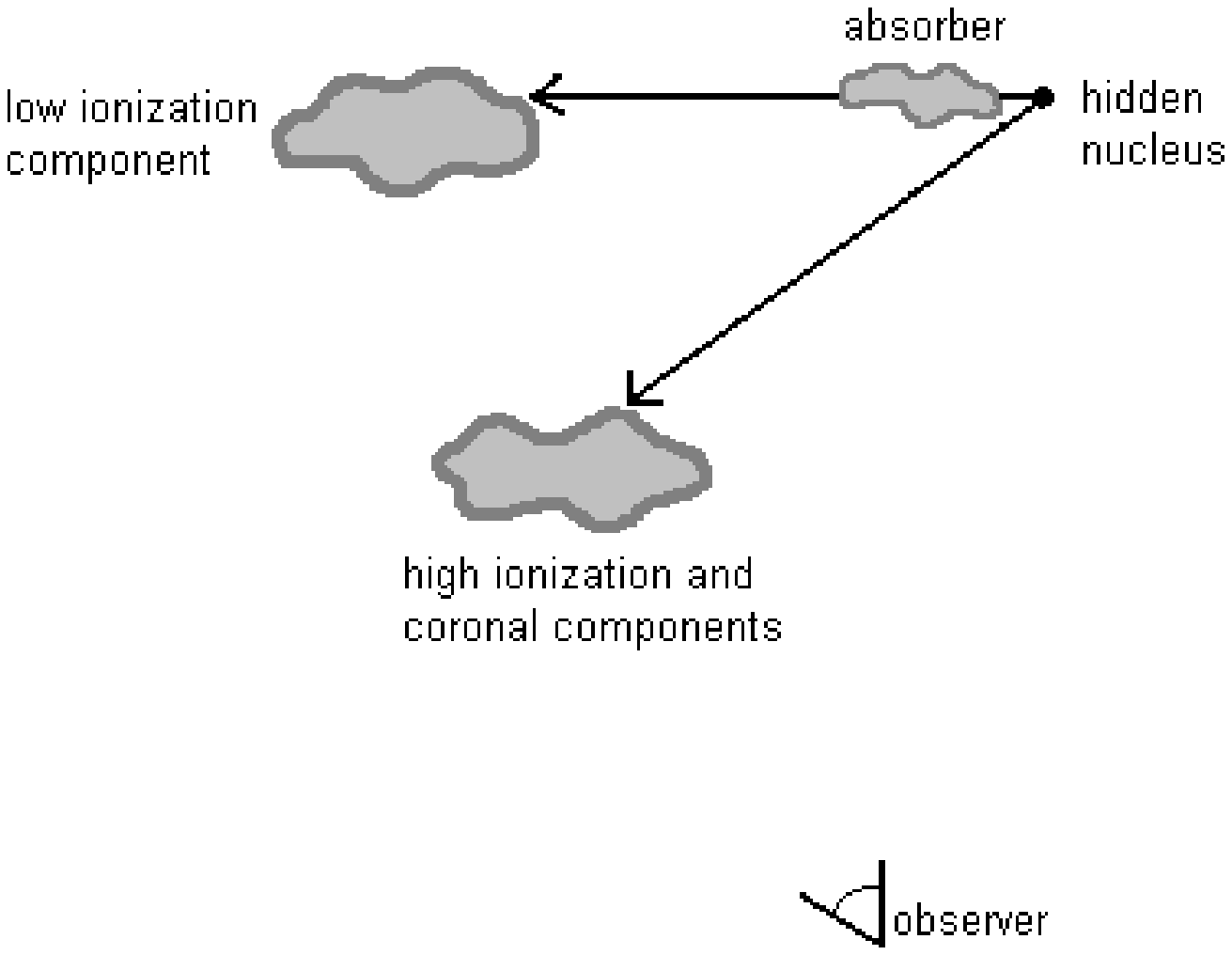]{Diagram showing the proposed geometry of the
inner NLR of NGC 1068. 
}\label{fig4} 

\figcaption[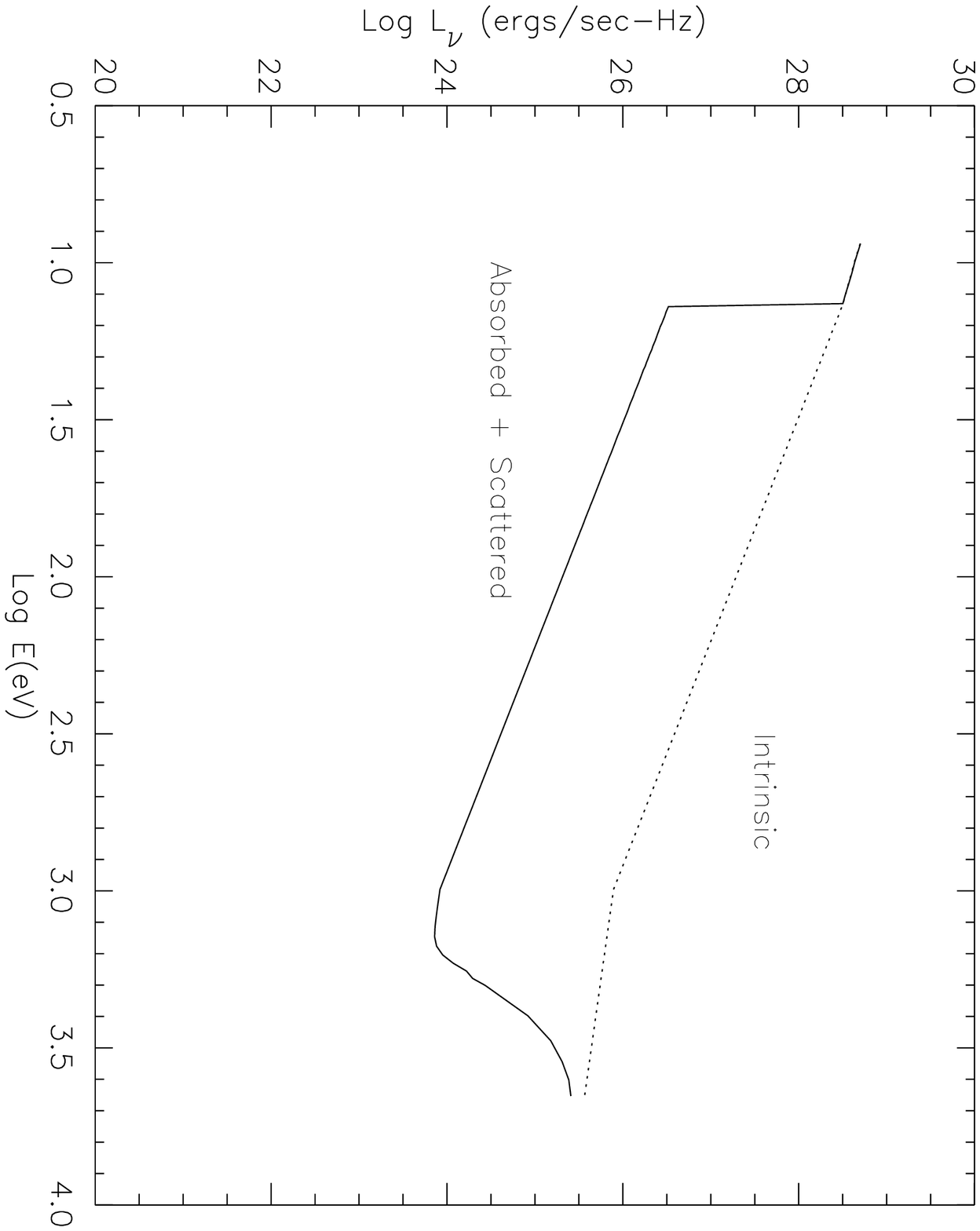]{The ionizing continuum used for the models.
The solid line is the absorbed $+$ scattered continuum used for
LOWION; the dotted line is the unabsorbed continuum used for
HIGHION and CORONAL.
}\label{fig5} 
 
\clearpage
\begin{deluxetable}{lllll}
\tablecolumns{5}
\footnotesize
\tablecaption{NGC 1068 Emission-Line Redshifts and Line Ratios
(relative to H$\beta$$^{a}$)\label{tbl-1}}
\tablewidth{0pt}
\tablehead{
\colhead{} & \colhead{c$z$$^{b}$} & 
\colhead{Ionization} &
\colhead{Observed} & 
\colhead{Reddening} \\
\colhead{} & \colhead{km s$^{-1}$} & 
\colhead{Energy (eV)$^{c}$} &
\colhead{Ratio} & 
\colhead{Corrected$^{d}$} 
}
\startdata
L$\alpha$ $\lambda$1216  & 1455   & 13.6    &11.54 $\pm$ 1.04 
&30.75 $\pm$ 5.65 \\
N V $\lambda$1240  & 1524   & 77.4  & 6.28 $\pm$ 0.50 &16.19 $\pm$ 2.86 \\
O I $\lambda$1302  & 2028   &       & 0.36 $\pm$ 0.04 & 0.84 $\pm$ 0.14 \\
C II $\lambda$1335 & 2316   & 11.3  & 0.43 $\pm$ 0.12 & 0.98 $\pm$ 0.19 \\
O IV] $\lambda$1402 $+$ (Si IV $\lambda$1398)  & 1713 & 77.4 & 2.34 $\pm$ 
0.29 & 4.99 $\pm$ 0.74 \\
$[$Si VIII] $\lambda$1441 & 1251 &246.5 & 0.25 $\pm$ 0.07 & 0.51 $\pm$ 0.10 \\
N IV] $\lambda$1486 &  969 & 47.4 & 0.38 $\pm$0.08 & 0.76 $\pm$ 0.13 \\
C IV $\lambda$1550  & 1530  & 47.9 &10.02 $\pm$ 0.70 &19.83 $\pm$ 2.53 \\ 
He II $\lambda$1640 &  915  & 54.4 & 2.25 $\pm$ 0.25 & 4.34 $\pm$ 0.57\\
C III] $\lambda$1909 $+$ (Si III] $\lambda\lambda$1883,1892) &  771
& 24.4 & 3.52 $\pm$ 0.30 & 7.16 $\pm$ 0.96 \\ 
$[$Si VII] $\lambda$2148 &  783 &205.1 & 0.29 $\pm$ 0.08 & 0.76 $\pm$ 0.15\\ 
C II] $\lambda$2326 ($+$ O III] $\lambda$2321) & 1509 & 11.3
              & 0.22 $\pm$ 0.06 & 0.47 $\pm$ 0.09 \\
$[$Ne IV] $\lambda$2423 & 1338 & 63.5 & 0.76 $\pm$ 0.11 & 1.44 $\pm$ 0.20 \\
$[$Fe XI] $\lambda$2649 &  531 &262.1 & 0.49 $\pm$ 0.11 & 0.77 $\pm$ 0.13 \\
Mg II $\lambda$2800 & 1767 &  7.6 & 1.32 $\pm$ 0.17 & 1.91 $\pm$ 0.21 \\
O III $\lambda$3133 & 1167 & 35.1 & 0.87 $\pm$ 0.15 & 1.13 $\pm$ 0.16 \\
He II $\lambda$3204 & 1452 & 54.4 & 0.70 $\pm$ 0.20 & 0.89 $\pm$ 0.21 \\
$[$Ne V] $\lambda$3346 & 1167 & 97.1 & 1.41 $\pm$ 0.15 & 1.74 $\pm$ 0.17 \\
$[$Ne V] $\lambda$3426 & 1377 & 97.1 & 4.05 $\pm$ 0.31 & 4.94 $\pm$ 0.36 \\
$[$Fe VII] $\lambda$3588 & 1155 &100.0 & 0.39 $\pm$ 0.07 & 0.46 $\pm$ 0.07 \\
$[$O II] $\lambda$3727 & 1110 & 13.6 & 0.48 $\pm$ 0.08 & 0.56 $\pm$ 0.08 \\
$[$Fe VII] $\lambda$3760 & 1356 &100.0 & 0.74 $\pm$ 0.07 & 0.85 $\pm$ 0.07 \\
\tablebreak
$[$Ne III] $\lambda$3869 & 1620 & 41.0 & 2.07 $\pm$ 0.18 & 2.35 $\pm$ 0.19 \\
$[$Ne III] $\lambda$3967 & 1542 & 41.0 & 0.77 $\pm$ 0.10 & 0.86 $\pm$ 0.10 \\
$[$S II] $\lambda$4072   & 1590 & 10.4 & 0.29 $\pm$ 0.05 & 0.33 $\pm$ 0.05 \\
H$\delta$ $\lambda$4101   & 1689 & 13.6 & 0.30 $\pm$ 0.05 & 0.33 $\pm$ 0.05 \\  
H$\gamma$ $\lambda$4341  & 1209 & 13.6 & 0.62 $\pm$ 0.06 & 0.66 $\pm$ 0.06\\
$[$O III] $\lambda$4363  & 1305 & 35.1 & 0.40 $\pm$ 0.05 & 0.43 $\pm$ 0.05\\
He II $\lambda$4686      & 1236 & 54.4 & 0.59 $\pm$ 0.05 & 0.60 $\pm$ 0.05\\ 
H$\beta$ $\lambda$4861   & 1320    & 13.6 & 1.00 & 1.00 \\ 
$[$O III] $\lambda$4959  & 1173 & 35.1 & 5.02 $\pm$ 0.38 & 4.96 $\pm$ 0.38 \\
$[$O III] $\lambda$5007  & 1215 & 35.1 &15.39 $\pm$ 0.98 &15.12 $\pm$ 0.98 \\
$[$Ni XIII] $\lambda$5116 ? &  705 &352.0 & 0.06 $\pm$ 0.02 & 0.06 $\pm$ 
0.02 \\
$[$Fe VII] $\lambda$5159 &  999 &100.0 & 0.40 $\pm$ 0.04 & 0.38 $\pm$ 0.04 \\
$[$Fe VI] $\lambda$5176 & 1755 & 75.5 & 0.22 $\pm$ 0.03 & 0.21 $\pm$ 0.03 \\
$[$Fe XIV] $\lambda$5303 &  708 &361.0 & 0.82 $\pm$ 0.06 & 0.78 $\pm$ 0.06 \\
$[$Ar X] $\lambda$5539 ? &  672 &422.6 & 0.15 $\pm$ 0.03 & 0.14 $\pm$ 0.03 \\
$[$Fe VII] $\lambda$5721 & 1092 &100.0 & 0.92 $\pm$ 0.07 & 0.83 $\pm$ 0.07 \\
He I $\lambda$5876 & 1653 & 24.6 & 0.29 $\pm$ 0.12 & 0.25 $\pm$ 0.12 \\
$[$Fe VII] $\lambda$6087 & 1113 &100.0 & 1.25 $\pm$ 0.10 & 1.08 $\pm$ 0.10 \\
$[$O I] $\lambda$6300 ($+$ [S III] $\lambda$6312) &  720 & -- 
              & 0.32 $\pm$ 0.03 & 0.27 $\pm$ 0.03 \\
$[$Fe X] $\lambda$6374 ($+$ [O I] $\lambda$6364)  &  720 &235.0 
              & 0.95 $\pm$ 0.07      & 0.80 $\pm$ 0.07     \\
$[$N II] $\lambda$6548 & 1200 & 14.5       
& 1.18 $\pm$ 0.22               & 0.98 $\pm$ 0.22     \\
H$\alpha$ $\lambda$6563 & 1167 & 13.6 
              & 3.39 $\pm$ 0.51   & 2.81 $\pm$ 0.51      \\
$[$N II] $\lambda$6584  & 1197 & 14.5
              & 3.54 $\pm$ 0.65                         & 2.94 $\pm$ 0.66     \\
\tablebreak
$[$Ni XV] $\lambda$6702 ? &  618 &430.0 & 0.07 $\pm$ 0.02 & 0.06 $\pm$ 0.02 \\
$[$S II] $\lambda$6716 & 1233  & 10.4
              & 0.21 $\pm$ 0.03                         & 0.17 $\pm$ 0.03      \\
$[$S II] $\lambda$6731 & 1230  & 10.4
              & 0.26 $\pm$ 0.03                         & 0.21 $\pm$ 0.04      \\
$[$Ar V] $\lambda$7005 ? & 1422 & 59.8 & 0.18 $\pm$ 0.03 & 0.15 $\pm$ 0.03 \\
He I $\lambda$7065 & 1014 & 24.6 & 0.10 $\pm$ 0.02 & 0.08 $\pm$ 0.02 \\
$[$Ar III] $\lambda$7136 & 1455 & 27.6 & 0.37 $\pm$ 0.04 & 0.29 $\pm$ 0.05 \\
$[$O II] $\lambda$7325 & 1353 & 13.6
              & 0.30 $\pm$ 0.04                    & 0.24 $\pm$ 0.04      \\
$[$S XII] $\lambda$7611 &  612 &504.7 & 0.47 $\pm$ 0.05 & 0.36 $\pm$ 0.05 \\
$[$Ar III] $\lambda$7751 & 1590 & 27.6 & 0.09 $\pm$ 0.02 & 0.07 $\pm$ 0.02 \\
$[$Fe XI] $\lambda$7892  &  654 &262.1 & 0.72 $\pm$ 0.06 & 0.55 $\pm$ 0.06 \\
$[$S III] $\lambda$9069	 & 1257 & 23.3
              & 0.72 $\pm$ 0.09      & 0.51 $\pm$ 0.09      \\
$[$S III] $\lambda$9532	 & 1116 & 23.3
              & 1.82 $\pm$ 0.15     & 1.28 $\pm$ 0.17      \\
& & & & \\

\tablenotetext{a}{Flux$_{H\beta}$ $=$ 18.07 
($\pm$ 1.11) x 10$^{-15}$ ergs cm$^{-2}$ s$^{-1}$.}
\tablenotetext{b}{c$z$ $=$ 1140 km s$^{-1}$ for neutral hydrogen in the host galaxy.}
\tablenotetext{c}{ionization energies from Allen (1968); for blended
lines, the ionization energy is for the strongest component.}
\tablenotetext{d}{E$_{B-V}$ $=$ 0.17 $\pm$ 0.03.}

\enddata
\end{deluxetable}
 
\clearpage
\begin{deluxetable}{lllll}
\tablecolumns{5}
\footnotesize
\tablecaption{Line Ratios from Model Components, Composite, and
Observations
(relative to H$\beta$)\label{tbl-2}}
\tablewidth{0pt}
\tablehead{
\colhead{} & \colhead{HIGHION$^{a}$} & \colhead{LOWION$^{b,c}$} & 
\colhead{Composite$^{d}$} &
\colhead{Observed$^{e}$}
}
\startdata
C III $\lambda$977                     & (1.47)             & (1.08)
              & (1.37)                         & -- \\            
N III $\lambda$990                     & (0.64)             & (0.78)
              & (0.68)                         & -- \\            
L$\alpha$ $\lambda$1216                     &34.18             &40.44 
              &35.74                         &30.17 $\pm$ 5.65 \\            
N V $\lambda$1240           	      	     & 3.33             & 0.00 
              & 2.50                         &16.19 $\pm$ 2.86 \\
C II $\lambda$1335                           & 0.03             & 0.09 (4.02)
              & 1.01                         & 0.98 $\pm$ 0.19 \\
O IV] $\lambda$1402 $+$ Si IV $\lambda$1398  & 4.74             & 0.00
              & 3.56                         & 4.99 $\pm$ 0.19 \\
N IV] $\lambda$1486         	      	     & 2.99             & 0.00 
              & 2.24                         & 0.76 $\pm$ 0.13 \\
C IV $\lambda$1550          	      	     &32.52             & 0.03 
              &24.40                         &19.83 $\pm$ 2.53   \\
He II $\lambda$1640         	      	     & 6.30             & 1.29
              & 5.05                         & 4.34 $\pm$ 0.57  \\
O III $\lambda$1663                          & 1.71             & 0.09
              & 1.31                         & --               \\
N III $\lambda$1750                          & 0.81             & 0.05
              & 0.62                         & --               \\
C III] $\lambda$1909 $+$ Si III] $\lambda\lambda$1883,1892   & 6.66 & 1.04 
              & 5.25                         & 7.16 $\pm$ 0.96   \\
C II] $\lambda$2326 $+$ O III] $\lambda$2321 & 0.24             & 1.96 
              & 0.67                         & 0.47 $\pm$ 0.09     \\
$[$Ne IV] $\lambda$2423     	      	     & 1.62             & 0.01 
              & 1.22                         & 1.44 $\pm$ 0.20     \\
$[$O II] $\lambda$2470      	      	     & 0.00             & 0.77
              & 0.19                         & --                   \\
Mg II $\lambda$2800         	      	     & 0.00             & 3.05 (8.54)
              & 2.13                         & 1.91 $\pm$ 0.21      \\
He II $\lambda$3204                   	     & 0.36             & 0.08 
              & 0.29                         & 0.89 $\pm$ 0.21      \\
$[$Ne V] $\lambda$3346      	      	     & 1.83             & 0.00
              & 1.38                         & 1.74 $\pm$ 0.17      \\
$[$Ne V] $\lambda$3426      	      	     & 4.99             & 0.00 
              & 3.74                         & 4.94 $\pm$ 0.36    \\
$[$Fe VII] $\lambda$3588    	      	     & 0.46             & 0.00 
              & 0.34                         & 0.44 $\pm$ 0.07  \\
$[$O II] $\lambda$3727      	      	     & 0.00             & 2.81 
              & 0.70                         & 0.56 $\pm$ 0.08  \\
$[$Fe VII] $\lambda$3760    	      	     & 0.63             & 0.00
              & 0.48                         & 0.85 $\pm$ 0.07    \\
$[$Ne III] $\lambda$3869                     & 1.12             & 1.67 
              & 1.26                         & 2.35 $\pm$ 0.19  \\
$[$Ne III] $\lambda$3967 + H$\epsilon$       & 0.50             & 0.68 
              & 0.55                         & 0.86 $\pm$ 0.10  \\
$[$S II] $\lambda$4072                	     & 0.00             & 0.87
              & 0.22                         & 0.33 $\pm$ 0.05  \\
H$\delta$ $\lambda$4100 	      	     & 0.26             & 0.26 
              & 0.26                         & 0.33 $\pm$ 0.05   \\
H$\gamma$ $\lambda$4340      	      	     & 0.47             & 0.47
              & 0.47                         & 0.66 $\pm$ 0.06   \\
\tablebreak
$[$O III] $\lambda$4363      	      	     & 0.63             & 0.07 
              & 0.49                         & 0.43 $\pm$ 0.05    \\
He II $\lambda$4686          	      	     & 0.87             & 0.19 
              & 0.70                         & 0.60 $\pm$ 0.05    \\
H$\beta$                 	      	     & 1.00 	           & 1.00 
             & 1.00    	     & 1.00   \\
$[$O III] $\lambda$4959                      & 6.60              & 2.85 
              & 5.66                         & 4.96 $\pm$ 0.38   \\
$[$O III] $\lambda$5007                      &19.80              & 8.56 
              &16.99                         &15.12 $\pm$ 0.98    \\
$[$Fe VII] $\lambda$5721     	      	     & 0.79              & 0.00
              & 0.60                         & 0.83 $\pm$ 0.07     \\
He I $\lambda$5876           	      	     & 0.02              & 0.13 
              & 0.05                         & 0.25 $\pm$ 0.12     \\
$[$Fe VII] $\lambda$6087     	      	     & 1.18              & 0.00 
              & 0.88                         & 1.08 $\pm$ 0.10      \\
$[$O I] $\lambda$6300 $+$ [S III] $\lambda$6312& 0.00              & 2.29
              & 0.57                         & 0.27 $\pm$ 0.03    \\
$[$O I] $\lambda$6364 $+$ $[$Fe X] $\lambda$6374  & 1.17           & 0.71 
              & 1.06                         & 0.80 $\pm$ 0.07     \\
$[$N II] $\lambda$6548         	             & 0.00              & 2.65
              & 0.66                         & 0.98 $\pm$ 0.22     \\
H$\alpha$ $\lambda$6563      	      	     & 2.78              & 2.94 
              & 2.82                         & 2.81 $\pm$ 0.51      \\
$[$N II] $\lambda$6584         	             & 0.00              & 7.65
              & 1.91                         & 2.94 $\pm$ 0.66     \\
$[$S II] $\lambda$6716	                      & 0.00              & 1.03
              & 0.26                         & 0.17 $\pm$ 0.03      \\
$[$S II] $\lambda$6731	                      & 0.00              & 1.22
              & 0.30                         & 0.21 $\pm$ 0.04      \\
$[$O II] $\lambda$7325	                      & 0.00              & 0.98
              & 0.24                         & 0.24 $\pm$ 0.04      \\
$[$S III] $\lambda$9069	                      & 0.00              & 1.17
              & 0.30                         & 0.51 $\pm$ 0.09      \\
$[$S III] $\lambda$9532	                      & 0.01              & 3.08
              & 0.72                         & 1.28 $\pm$ 0.17      \\
& & & & \\

\tablenotetext{a}{U = 10$^{-1.5}$, n$_{H}$=6x10$^{4}$ cm$^{-3}$,
N$_{H}$ = 1 x 10$^{21}$ cm$^{-2}$; Flux$_{H\beta}$ = 5.06 ergs cm$^{-2}$ s$^{-1}$,
emitting area $=$ 1.3 x 10$^{38}$ cm$^{2}$, F$_{c}$ $=$ 0.002.}
\tablenotetext{b}{U = 10$^{-3.2}$, N$_{H}$=3x10$^{4}$ cm$^{-3}$
N$_{H}$ = 1 x 10$^{21}$ cm$^{-2}$; Flux$_{H\beta}$ = 0.25 
ergs cm$^{-2}$ s$^{-1}$,
emitting area $=$ 8.8 x 10$^{38}$ cm$^{2}$, F$_{c}$ $=$ 0.012.}
\tablenotetext{c}{values in parenthesis from CLOUDY90, turbulent
velocity $=$ 50 km s$^{-1}$.}
\tablenotetext{d}{75\% HIGHION, 25\% LOWION.}
\tablenotetext{e}{dereddened; E$_{B-V}$ $=$ 0.17 $\pm$ 0.03.}

\enddata
\end{deluxetable}

\clearpage
\begin{deluxetable}{llllllllll}
\tablecolumns{10}
\footnotesize
\tablecaption{Predicted Mean Ionization Fractions (from CORONAL model)
\label{tbl-3}}
\tablewidth{0pt}
\tablehead{
\colhead{Element} & \colhead{VII} & 
\colhead{VIII} &
\colhead{IX} &
\colhead{X} &
\colhead{XI} &
\colhead{XII} &
\colhead{XIII} &
\colhead{XIV} &
\colhead{XV} 
}
\startdata
Si & 0.001$^{a}$ & 0.034$^{a}$ & 0.191 & 0.339 & 0.288 & 0.099 & 0.046 & 0.002 & -- \\
S  & -- & 0.019 & 0.124 & 0.235 & 0.261 & 0.221$^{a}$ & 0.107 & 0.025 & 0.007 \\
Ar & 0.001 & 0.012 & 0.101 & 0.179$^{a}$ & 0.229 & 0.228 & 0.150 & 0.077 & 0.019 \\
Fe & --$^{a}$ & -- & 0.003 & 0.020$^{a}$ & 0.076$^{a}$ & 0.188 & 0.252 
& 0.193$^{a}$ & 0.160 \\
Ni & -- & -- & 0.001 & 0.022 & 0.106 & 0.180 & 0.181$^{a}$ & 0.170 & 0.144$^{a}$ \\

\tablenotetext{a}{Observed in hot spot spectrum.}

\enddata
\end{deluxetable}

\clearpage
\plotone{fig1.eps}

\clearpage
\plotone{fig2.eps}

\clearpage
\plotone{fig3.eps}

\clearpage
\plotone{fig4.ps}

\clearpage
\plotone{fig5.eps}

\end{document}